\documentclass[runningheads]{llncs}
\usepackage{graphicx}

\usepackage{multirow}
\usepackage{tabu}
\usepackage{bbm}
\usepackage{diagbox}
\usepackage[english]{babel}
\usepackage{comment}
\usepackage{array}
\usepackage{amstext}
\usepackage{makecell}
\usepackage{caption}
\usepackage[perpage]{footmisc}
\usepackage{babel}
\usepackage{booktabs} 
\usepackage{color}
\usepackage{amsmath}
\usepackage[pagebackref=false,breaklinks=true,letterpaper=true,colorlinks,bookmarks=false]{hyperref}

\newcommand{\ie}{\textit{i.e.}}
\newcommand{\eg}{\textit{e.g.}}

\begin{document}
\title{RibSeg Dataset and Strong Point Cloud Baselines for Rib Segmentation from CT Scans}
\titlerunning{RibSeg: Rib Segmentation from CT Scans}

\author{Jiancheng Yang\inst{1,2,}\thanks{These authors have contributed equally: Jiancheng Yang and Shixuan Gu.}  \and Shixuan Gu\inst{1,\star} \and Donglai Wei\inst{3} \and \\Hanspeter Pfister\inst{3} \and Bingbing Ni\inst{1}\thanks{Corresponding author: Bingbing Ni (nibingbing@sjtu.edu.cn).}}

\authorrunning{J. Yang et al.}

\institute{Shanghai Jiao Tong University, Shanghai, China\\
\email{nibingbing@sjtu.edu.cn}	\\
	\and Dianei Technology, Shanghai, China
	\and Harvard University, Cambridge MA, USA
}

\maketitle            

\begin{abstract}
Manual rib inspections in computed tomography (CT) scans are clinically critical but labor-intensive, as 24 ribs are typically elongated and oblique in 3D volumes. Automatic rib segmentation methods can speed up the process through rib measurement and visualization. However, prior arts mostly use in-house labeled datasets that are publicly unavailable and work on dense 3D volumes that are computationally inefficient. To address these issues, we develop a labeled rib segmentation benchmark, named \emph{RibSeg}, including 490 CT scans (11,719 individual ribs) from a public dataset. For ground truth generation, we used existing morphology-based algorithms and manually refined its results. Then, considering the sparsity of ribs in 3D volumes, we thresholded and sampled sparse voxels from the input and designed a point cloud-based baseline method for rib segmentation. The proposed method achieves state-of-the-art segmentation performance (Dice~$\approx95\%$) with significant efficiency ($10\sim40\times$ faster than prior arts). The RibSeg dataset, code, and model in PyTorch are available at \url{https://github.com/M3DV/RibSeg}.

\keywords{rib segmentation \and rib centerline \and medical image dataset \and point clouds \and computed tomography.}
\end{abstract}

\section{Introduction}

\begin{figure}[tb]
    \centering
	\includegraphics[width=\linewidth]{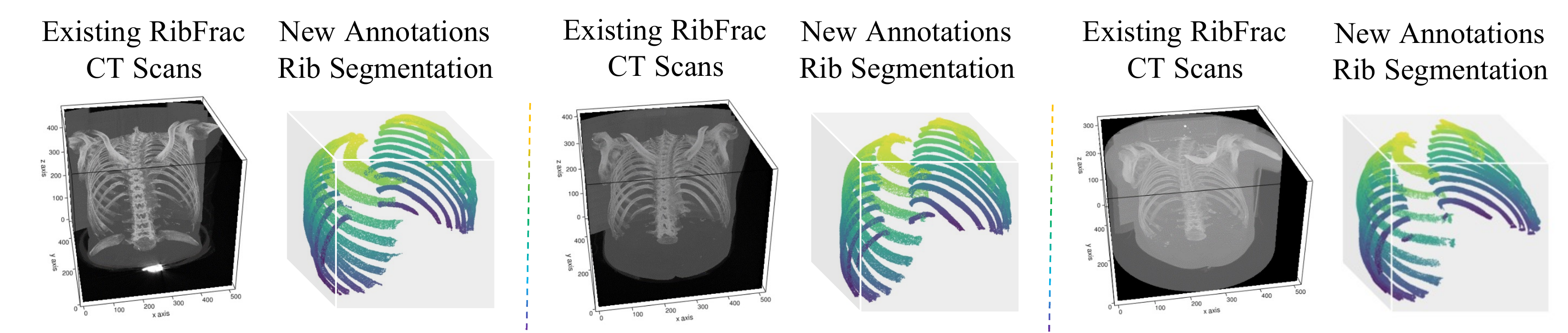}
	\caption{\textbf{Illustration of RibSeg Dataset}: 490 CT scans from the existing RibFrac dataset~\cite{jin2020deep} and new annotations for rib segmentation. Colors in rib segmentation figures denotes the axial depth.}
	\label{fig:RibSeg_dataset}
\end{figure}

Detection and diagnosis of rib-related diseases, \eg, rib fracture and bone lesions, is essential and common in clinical practice and forensics. For instance, rib fracture detection can identify chest trauma severity that accounts for $10\%\sim15\%$ of all traumatic injuries~\cite{sirmali2003comprehensive}, and bone metastases are common in solid tumours~\cite{coleman2014bone}. Chest computed tomography (CT) is a primary choice for examining chest trauma thanks to its advantage in revealing occult fractures and fracture-related complications~\cite{jin2018low}. However, understanding ribs in CT imaging is challenging: 24 ribs in human bodies are typically elongated and oblique in 3D volumes, with complex geometries across numerous CT sections; in other words, a large number of CT slices must be evaluated sequentially, rib-by-rib and side-by-side, which is tedious and labor-intensive for clinicians. Besides, fractures/lesions could be inconspicuous when reading 2D slices. These challenges urge the development and application of rib visualization tools, \eg, rib unfolding~\cite{ringl2015ribs}, whose core technique is (automatic) rib segmentation and centerline extraction.

A few prior arts have addressed rib segmentation~\cite{shen2004tracing,staal2007automatic,klinder2007automated} before the era of deep learning. Rib tracing is a popular method, while it is highly sensitive to initially detected seed points and vulnerable to local ambiguities. Supervised deep learning-based segmentation~\cite{lenga2018deep} from CT volumes is robust as it learns hierarchical visual features from raw voxels; However, this study does not consider the sparsity and elongated geometry of ribs in 3D volumes. There are also studies focusing on rib centerline extraction instead of full rib segmentation, \eg, rib tracing~\cite{ramakrishnan2011automatic} and deformable template matching~\cite{wu2012learning}. This study focuses on rib segmentation, where rib centerlines could be extracted with geometric post-processing algorithms. 

Although researchers have made progress in rib segmentation and centerline extraction, there is no public dataset in this topic, making it difficult to benchmark existing methods and develop downstream applications (\eg, rib fracture detection). Besides, existing methods work on the dense 3D volumes instead of the sparse rib voxels, which are computationally inefficient: around $5s$ to $20s$ to segment ribs~\cite{lenga2018deep} or $40s$ to extract rib centerlines~\cite{wu2012learning}. To address these issues, we first develop a large-scale CT dataset, named \emph{RibSeg} for rib segmentation (Fig.~\ref{fig:RibSeg_dataset}). The CT scans come from the public RibFrac dataset~\cite{jin2020deep} consisting of 660 chest-abdomen CT scans for rib fracture segmentation, detection, and classification. As ribs are relatively recognizable compared to other anatomical structures, we use hand-crafted 3D image processing algorithms to generate segmentation and then manually check and refine the labels. This procedure, with little annotation effort, generates visually satisfactory rib segmentation labels for 490 CT scans. Moreover, considering the sparsity of ribs in 3D volumes ($<0.5\%$ voxels) and high HU values ($>200$) of bone structures in CT scans, we propose an efficient point cloud-based model to segment ribs, which is an order of magnitude faster than previous methods. The proposed method converts dense CT volume into sparse point clouds via thresholding and random downsampling, and produces high-quality and robust rib segmentation (Dice~$\approx 95\%$). We can further extract rib centerlines from the predicted rib segmentation with post-processing. 

The RibSeg dataset could be used for the development of downstream rib-related applications. Besides, considering the differences from standard medical image datasets~\cite{antonelli2021medical,yang2020medmnist} with pixel/voxel grids, the elongated shapes and oblique poses of ribs enable the RibSeg dataset to serve as a benchmark for curvilinear structures and geometric deep learning (\eg, point clouds).

\paragraph{\textbf{Contributions.}} 1) The first public benchmark for rib segmentation, which enables downstream applications and method comparison. 2) A novel point-based perspective on modeling 3D medical images beyond voxel grids. 3) A point cloud-based rib segmentation baseline with high efficiency and accuracy.

\section{Materials and Methods}
\subsection{RibSeg Dataset} \label{sec:dataset}

\paragraph{\textbf{Dataset Overview.}}
The RibSeg dataset uses the public computed tomography (CT) scans from RibFrac dataset~\cite{jin2020deep}, an open dataset with 660 chest-abdomen CT scans for rib fracture segmentation, detection, and classification. The CT scans are saved in NIFTI (.nii) format with volume sizes of $N \times 512 \times 512$, where $512 \times 512$ is the size of CT slices, and
$N$ is the number of CT slices (typically $300\sim400$). Most cases are confirmed with complete rib cages and manually annotated with at least one rib fracture by radiologists.

As ribs are relatively recognizable compared to other anatomical structures, we use a semi-automatic approach (see details in the following section) to generate rib segmentation, with hand-crafted morphology-based image processing algorithms, as well as manual checking and refinement. Though computationally intensive, this approach produces visually satisfactory labels with few annotation efforts. Finally, there are 490 qualified CT cases in the RibSeg dataset with 11,648 individual ribs in total, where each case has segmentation labels of 24 (or 22 in some cases) ribs. We also provide the rib centerline ground truth extracted from the rib segmentation labels. Note that the rib segmentation and centerline ground truth are imperfect, as the annotations are generated with algorithms. Besides, only voxels with higher HU than a threshold (200) are included, making the rib segmentation annotations in the RibSeg dataset hollow. However, we manually check the ground truth labels for rib segmentation and centerlines to ensure that the included 490 datasets are high-quality enough to develop downstream applications. The data split of the RibSeg dataset is summarized in Table~\ref{tab:data-overview}: training set (320 cases, to train the deep learning system), development set (a.k.a validation set, 50 cases, to tune hyperparameters of the deep learning system), and test set (120 cases, to evaluate the model). The RibSeg training, development, and test set are from those of the RibFrac dataset respectively, enabling the development of downstream applications (\eg, rib fracture detection) in the MICCAI 2020 RibFrac challenge\footnote{\url{https://ribfrac.grand-challenge.org/}}.

\begin{table}[tb]
	\caption{Overview of RibSeg dataest.} \label{tab:data-overview}
	\centering
	\begin{tabular*}{\hsize}{@{}@{\extracolsep{\fill}}lcc@{}}
		\toprule
		Subset & No. of CT Scans &  No. of Individual Ribs \\
		\midrule
		Training / Development / Test & 320 / 50 / 120 & 7,670 / 1,187 / 2,862 \\
		
		\bottomrule
	\end{tabular*}
	
\end{table}

\paragraph{\textbf{Semi-Automatic Annotation and Quality Control.}}
We describe the primary steps of the annotation procedure as follows:

\textit{Rib Segmentation.} For each volume, we first filter out non-target voxels by thresholding and removing regions outside the bodies. Considering the geometric differences between ribs and vertebra, we separate the ribs from vertebra using morphology-based image processing algorithms(\eg, dilation, erosion). In some cases, the segmentation result contains parts of the clavicle and scapula. Therefore, we manually locate those non-target voxels and remove them according to the coordinates of their connected components.

\textit{Centerline Extraction.} Based on the rib segmentation, we extract the
centerline by implementing the following procedure on each rib (connected component): randomly select two points at both ends of the cylinder dilated from the rib, calculate the shortest path between the points\footnote{\url{https://github.com/pangyuteng/simple-centerline-extraction}}~\cite{teng2011automated}, and smoothen the path to obtain centerline. This procedure produces high-quality centerlines even from coarse rib segmentation. At the end of extraction, we label both centerlines and rib segmentation in the order of top to bottom and left to right.

\textit{Manual Checking and Refinement.} The abnormal cases, along with the pursuit of high annotation quality, motivate us to perform laborious checking and refinement after both rib segmentation and centerline extraction stages. For instance, a few cases miss floating ribs after segmentation, which reduces the connected components in annotation to 22 or less. Hence we have to check and refine the annotation case by case manually. To recover and annotate missed ribs, we turn back to the previous stage to ensure segmentation completeness by modifying the corresponding components.

\subsection{Rib Segmentation from a Viewpoint of Point Clouds} \label{sec:method}

The key insight of the proposed method is that simple algorithm (\ie, thresholding in this study) can produce the candidate voxels for bone structures. Thus, we can avoid heavy computation on dense voxels with sparse point clouds instead. Besides, the point cloud methods use geometric information directly, reducing the texture bias of pixel/voxel-based CNNs~\cite{geirhos2018imagenet}. The point cloud viewpoint has the potential to generalize to other anatomical structures whose coarse prediction could be obtained cheaply.

\begin{figure}[tb]
    \centering
	\includegraphics[width=\linewidth]{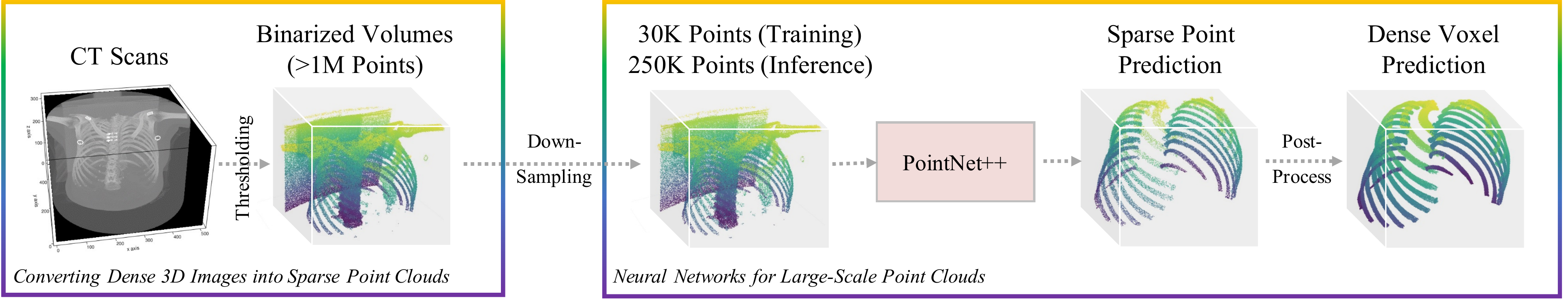}
	\caption{\textbf{Rib Segmentation from a Viewpoint of Point Clouds}. The CT volumes are first binarized to obtain candidate bone voxels as inputs, then a point cloud neural network (\eg, PointNet++~\cite{qi2017pointnet++}) is used to segment each point in downsampled input point clouds. Note that the downsampling scale is different during training (30K points) and inference (250K points). }
	\label{fig:model}
\end{figure}

\paragraph{\textbf{Deep Learning for Point Cloud Analysis.}} Deep learning for point cloud analysis~\cite{guo2020deep} has been an emerging field in computer vision thanks to the popularity of 3D sensors (\eg, LiDAR and RGB-D cameras) and computer graphics applications (\eg, games, VR/AR/MR), PointNet~\cite{qi2017pointnet} and DeepSet~\cite{zaheer2017deep} pioneer this direction, where a symmetric function (\eg, shared FC) is used for learning high-level representation before aggregation (\eg, pooling). Following studies introduces sophisticated feature aggregation based on spatial graphs~\cite{qi2017pointnet++,liu2019dynamic} or attention~\cite{yang2019modeling}. However, only a few studies have applied deep shape analysis in medical imaging scenarios~\cite{yang2020intra,wickramasinghe2020voxel2mesh,Yang2019ProbabilisticRA}. 

\paragraph{\textbf{Model Pipeline.}}
Considering the sparsity of ribs in 3D volumes ($<0.5\%$ voxels) and high HU values of bones in CT scans, we design a point cloud-based model to segment ribs on binarized sparse voxels. As depicted in Fig.~\ref{fig:model}, we first set a threshold of 200 HU (Hounsfield Unit) to filter out the non-bone voxels roughly. The resultant binarized volumes are randomly downsampled and converted to point sets for ease of computation before forwarding to the network. 

Our point cloud-based model is expected to infer dense predictions from large-scale point sets, which has to address the memory issue. Hence a custom PointNet++~\cite{qi2017pointnet++}, with its adjustable memory footprint, is adopted as backbone. Capable of learning local features with increasing contextual scales, PointNet++ has shown compelling robustness on sparse 3D point cloud segmentation tasks. Through set abstraction, geometric features of ribs can be extracted from binarized sparse voxels facilitating the rib segmentation task. For post-processing, the model output point prediction is converted back to volumes (voxel prediction) by morphology-based image processing algorithms.

\paragraph{\textbf{Model Training and Inference.}} During the training stage, batches are downsampled to 30,000 points per volume considering the trade-off between batch size and input size. We apply online data augmentations, including scaling, translation, and jittering, to all downsampled point sets before forwarding them to the neural network. The Adam optimizer is adopted to train all models end-to-end for 250 epochs with the batch size of 8 and cross-entropy loss (CE) as the loss function. The initial learning rate was set at 0.001 and decayed by a factor of 0.5 every 20 epochs with the lower bound of $10^{-5}$.

During the inference stage, volumes are converted to point sets with the size of 250,000. The model then produces dense point predictions on rib segmentation. The point predictions are converted back to dense volumes by dilation, and we obtain the voxel predictions of rib segmentation by taking the intersections between the dilated volumes and the binarized volumes.

\paragraph{\textbf{Model Evaluation.}} As the model only outputs sparse point predictions, we post-process the point predictions back to volumes (dense voxel predictions). Both point-wise and voxel-wise segmentation performance is evaluated,
\begin{equation}
Dice^{(L)}=2\cdot|y_{(L)}\cdot \hat{y}_{(L)}|/(|y_{(L)}|+| \hat{y}_{(L)}|),~ L \in \{P,V\},
\end{equation}
where $Dice^{(P,V)}$ indicates the sparse point-wise and dense voxel-wise Dice.

Apart from segmentation performance, we also report the missing ratio of individual ribs to evaluate the clinical applicability. Specifically, a missing of an individual rib $i$ is counted if $\textit{recall}_i<0.5$, and then the missing ratio can be calculated with ease. As the segmentation of first and twelfth rib pairs tend to be more difficult, we calculate and report the missing ratio of all/first/intermediate/twelfth rib pairs, as depicted in Table~\ref{tab:metrics}.

\section{Results}

\subsection{Quantitative Analysis} \label{sec:result}

For model accuracy comparison, we first implement a 3D UNet~\cite{cciccek20163d} taking patches of CT volumes as input with the same setting of FracNet~\cite{jin2020deep}. Moreover, we train two models with and without data augmentation, respectively. The models are evaluated with two input (point sets) sizes: 30K (input size in the training stage) and 250K. As point-wise Dice is only a proxy metric, we focus on voxel-wise Dice, as it is fair for any methods in rib segmentation. As depicted in Table.~\ref{tab:metrics}, all point-based methods significantly outperform voxel-based 3D UNet; Besides, as point-based methods take whole volumes as inputs, it is more efficient than voxel-based method. The methods with data augmentation are at least 2\% higher than the methods without data augmentation, and methods with large-scale input enjoy 0.9\%$\sim$1.3\% higher values, as dense point prediction leads to rich details in voxel prediction. When it comes to the missing ratio of ribs, the method with data augmentation and a large input size performs best. The comparison results show that training-time data augmentation and inference-time large input volume size can improve the result. These quantitative metrics also indicate the potentials of our method in clinical applications.

In terms of the run-time, point-based methods have a clear advantage. While methods with a 250K-point input size have a little bit higher time consumption (0.8s), it is acceptable in consideration of its performance boost. The inference time was measured with the implementation of PyTorch 1.7.1~\cite{Paszke2017AutomaticDI} and Python 3.7, on a machine with a single NVIDIA Tesla P100 GPU with Intel(R) Xeon(R) CPU @ 2.20 GHz and 150 G memory. As a reference, prior art~\cite{lenga2018deep} based on 3D networks takes a model forward time of 5$\sim$20 seconds, with a Dice value of $84\%$ at best. The previous work~\cite{wu2012learning} takes 40 seconds to extract the rib centerline. However, a direct comparison of metrics and speed is unfair since these results were measured with different infrastructures on different datasets. Note that, post-processing time is not included as it heavily depends on the implementation (\eg, programming languages, parallel computing).

\begin{table}[tb]
	\caption{Quantitative metrics on RibSeg test set, including Dice over sparse points ($Dice^{(P)}$), Dice over dense voxels after post-processing ($Dice^{(v)}$), ratio of missed all/first/intermediate/twelfth rib pairs (A/F/I/T) at recall$>0.5$, and the model forward time in second. Post-processing time is not included as it heavily depends on the implementation.}\label{tab:metrics}
	\centering
	\begin{tabular*}{\hsize}{@{}@{\extracolsep{\fill}}l|c|c|c|c@{}}
		\toprule
		Methods & $Dice^{(P)}$ & $Dice^{(V)}$ & Missed Ribs (A/F/I/T) & Forward (s)\\
		\midrule
		Voxel-Based 3D UNet~\cite{cciccek20163d,jin2020deep}&-& 86.3\% & 4.6\%/7.9\%/2.3\%/24.6\% & 30.63\\
		\midrule
		PN++~\cite{qi2017pointnet++} (30K) & 92.3\% & 91.0\% & 1.6\%/2.9\%/0.7\%/10.4\% & \bf 0.32 \\
		PN++~\cite{qi2017pointnet++} (250K) & 91.5\% & 92.3\% & 0.9\%/3.3\%/0.3\%/4.7\% & 1.12 \\
		PN++~\cite{qi2017pointnet++} (30K) + aug. & \bf94.9\% & 94.3\% & 1.1\%/0.8\%/0.4\%/9.0\% & \bf0.32 \\
		PN++~\cite{qi2017pointnet++} (250K) + aug. & 94.6\% & \bf 95.2\% & \bf0.6\%/0.4\%/0.2\%/5.2\% & 1.12 \\
		
		\bottomrule
	\end{tabular*}
\end{table}

\subsection{Qualitative Analysis}

\paragraph{\textbf{Visualization on Predicted Rib Segmentation.}}

\begin{figure}[tb]
    \centering
	\includegraphics[width=0.8\linewidth]{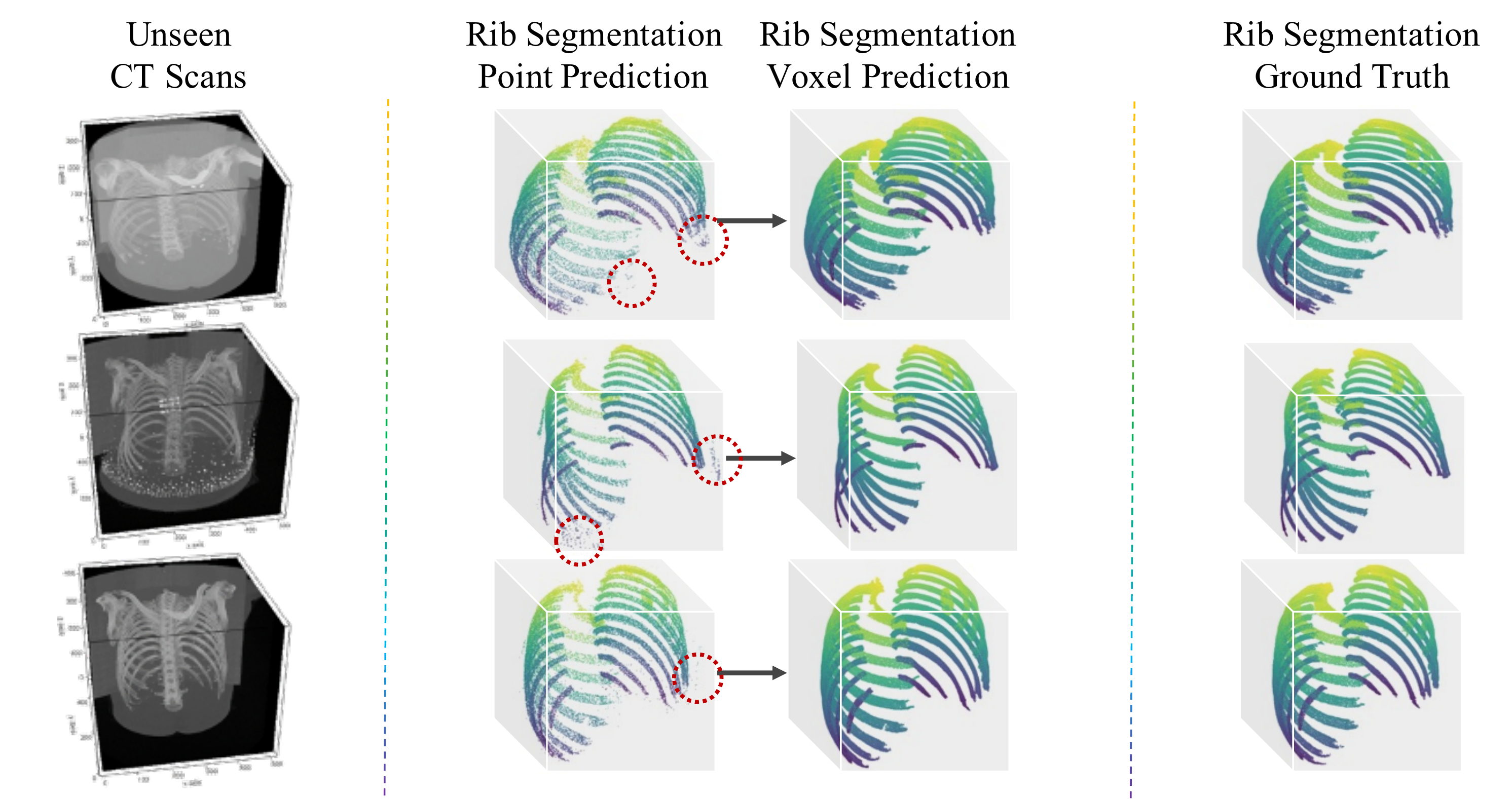}
	\caption{\textbf{Visualization of Predicted Rib Segmentation.}. Red circles denote imperfect (sparse) point prediction.} \label{fig:visualization_main}
\end{figure}

Fig.~\ref{fig:visualization_main} visualizes the point-level and voxel-level rib segmentation prediction. As depicted in Fig.~\ref{fig:visualization_main}, point predictions of the first 3 CT scans are visually acceptable, which can smoothly produce voxel-level segmentation predictions. The results are promising, and it is even hard to tell the visual difference between predictions and ground truths. 

As depicted in Fig.~\ref{fig:visualization_main}, the point prediction is imperfect even with high segmentation Dice. After manually-tuned post-processing, these imperfect predictions could be fixed to some degree. For instance, the point prediction of the second CT scan contains a certain part of the scapula that can be nicely filtered during post-processing. However, the point prediction of the third CT scan suffers a missing on the first pair of ribs, which can not be ignored. The post-process is not able to fix it and produces incomplete voxel segmentation. Despite the small missing part, the rib cages in the predictions on rib segmentation are still visually acceptable.

\begin{figure}[tb]
    \centering
	\includegraphics[width=\linewidth]{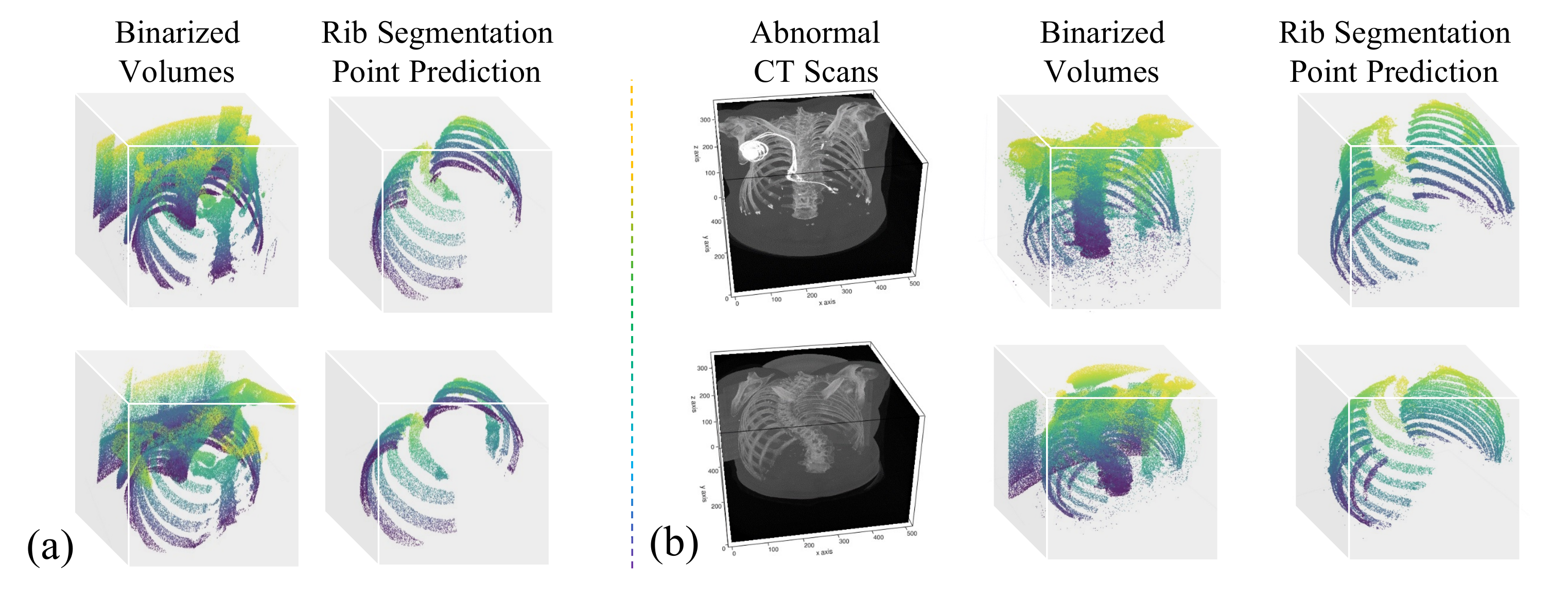}
	\caption{\textbf{Robustness Test on Extreme Cases.} (a) Point prediction on unseen incomplete rib cages. (b) Point prediction on unseen abnormal CT scans. }
	\label{fig:visualization_extreme}
\end{figure}

\paragraph{\textbf{Robustness Test on Extreme Cases.}}
To evaluate the robustness of the proposed point cloud-based model, three kinds of extreme cases are selected for inference: CT scans of incomplete rib cages, CT scans of serious spinal pathology, and CT scans containing metal objects inseparable from ribs (\eg, pacemaker). Incomplete CT scans are rather common in clinical cases, which makes the tests practically critical. As depicted in Fig.~\ref{fig:visualization_extreme}~(a), we randomly select unseen CT scans and take the upper half of their binarized volumes for inference. Delightfully, the point predictions are visually qualified for clinical applications.

For further robustness evaluation, we test on a case of serious spinal pathology and a case containing a pacemaker of high density. Regarding the case with a pacemaker, it is inseparable from ribs, which makes the segmentation extremely laborious to obtain. Hence we save the trouble of manual segmentation by setting it as an abnormal case when building the RibSeg Dataset. As depicted in Fig.~\ref{fig:visualization_extreme}~(b), our prediction on the case of spinal pathology looks complete as if it is nicely segmented. While the prediction on the other case contains a small number of voxels belong to the pacemaker, the whole rib cage is well segmented. Considering that all cases selected for the robustness test are unseen and geometrically difficult to segment, such promising prediction results may confirm the strong robustness of the proposed method as well as its potentials to be clinically applicable.

\begin{figure}[tb]
    \centering
	\includegraphics[width=\linewidth]{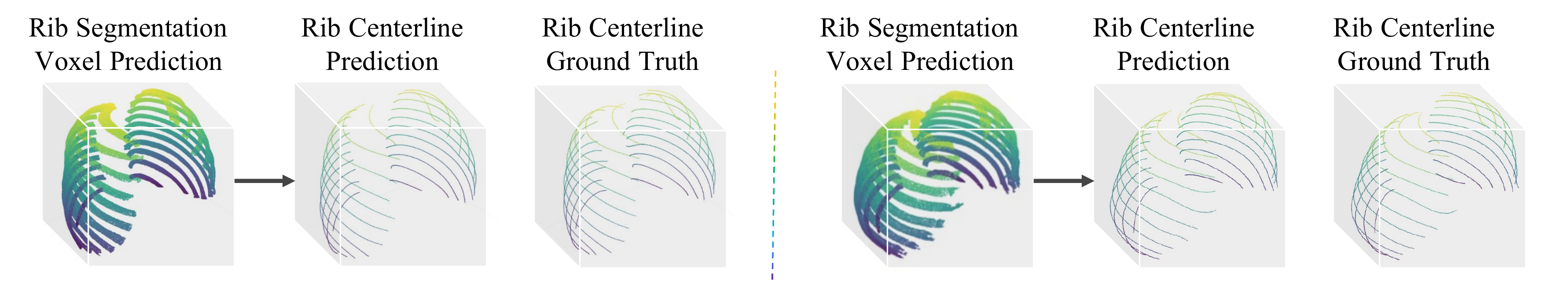}
	\caption{\textbf{Post-Processing Rib Centerlines} from predicted rib segmentation. }
	\label{fig:visualization_cl}
\end{figure}

\paragraph{\textbf{Post-Processing Rib Centerlines.}}
With the high-quality rib segmentation, as illustrated in Fig.~\ref{fig:visualization_cl}, the rib centerlines could be obtained by post-processing (\ie, shortest path between points, same as the procedure with ground truth in Sec.~\ref{sec:dataset}). Although not end-to-end, the rib centerline predictions are visually acceptable for most cases. However, the post-processing algorithms for rib centerlines are sensitive to rib fractures and other abnormal cases. Considering the high clinical importance of rib centerlines, it urges a more robust method for rib centerline extraction with rib segmentation and centerline labels provided by the RibSeg dataset.

\section{Conclusion and Further Work}
We built the RibSeg dataset, which is the first open dataset for rib segmentation. On this dataset, we benchmarked a point cloud-based method with high performance and significant efficiency. The proposed method shows potentials to be clinically applicable, enhancing the efficiency and performance of downstream tasks, such as the diagnosis of rib fractures and bone lesions. Besides the clinical application, the RibSeg dataset could also serve as an interesting benchmark for curvilinear structures and geometric deep learning (\eg, point clouds), considering the special geometry of rib structures.

There are several limitations in this study. The annotations in this paper are generated with hand-crafted morphological algorithms, and then manually checked by a junior radiologist with 3D Slicers. 
While such pipeline reduces the annotation cost, it cannot handle cases when the initial automatic method fails. Thus, we only managed to annotate the segmentation for 490 cases out of the 660 cases in the RibFrac dataset.
Also, for the centerline extraction task that is essential for rib-related applications, we take a two-stage approach and apply heavy post-processing method to the first-stage segmentation result.
Such approach is sensitive to rib fractures and segmentation errors for other abnormal cases and a more robust method will be favorable.

\subsubsection{Acknowledgment.}
This work was supported by the National Science Foundation of China  (U20B2072, 61976137).

\bibliographystyle{splncs04}
\bibliography{reference}

\end{document}